\documentclass[12pt]{iopart}
 \usepackage{graphicx,amsfonts,color,amssymb,latexsym,amsbsy}

 \newcommand\Zb{\mathbb{Z}}

 \newcommand{\ben}{\begin{equation*}}
 \newcommand{\ebn}{\end{equation*}}
 \newcommand\be{\begin{equation}}
 \newcommand\eb{\end{equation}}
 
 \newcommand\Kr{\mathcal{K}}

\begin{document}
 \title{Finite-lattice form factors in free-fermion models}
 \author{N Iorgov$^1$, O Lisovyy$^2$}
 \address{
  $^1$ Bogolyubov Institute for Theoretical Physics,
 03680, Kyiv, Ukraine}
  \address{$^2$ Laboratoire de Math\'ematiques et Physique Th\'eorique CNRS/UMR 6083,
  Universit\'e de Tours, Parc de Grandmont, 37200 Tours, France}
  \eads{\mailto{iorgov@bitp.kiev.ua}, \mailto{lisovyi@lmpt.univ-tours.fr}}

 \begin{abstract}
 We consider the general $\mathbb{Z}_2$-symmetric free-fermion model on the finite periodic lattice,
 which includes as special cases the Ising model on the square and triangular lattices and
 $\mathbb{Z}_n$-symmetric BBS $\tau^{(2)}$-model with $n=2$.
 Translating Kaufman's fermionic approach to diagonalization of Ising-like transfer matrices
 into the language of Grassmann integrals, we
 determine the transfer matrix eigenvectors and observe that
 they coincide with the eigenvectors of
 a square lattice Ising transfer matrix. This allows to find
 exact finite-lattice form factors of spin operators for the statistical model and
 the associated finite-length quantum chains,
 of which the most general is equivalent to the XY chain in a transverse field.
 \end{abstract}


 \section{Introduction}
 Consider a two-dimensional $M\times N$ square lattice with spins $\sigma=\pm1$ living
 at each of its sites. The most general $\mathbb{Z}_2$-symmetric plaquette Boltzmann weight
 is given by
 \be\label{gbweight}
 W(\sigma_1,\sigma_2,\sigma_3,\sigma_4)=a_0\Bigl(1+\!\!\!\sum_{1\leq i<j\leq 4}a_{ij}\sigma_i\sigma_j
 +a_4\sigma_1\sigma_2\sigma_3\sigma_4\Bigr).
 \eb
 Partition function of the corresponding statistical model can be represented as an integral over four-component
 lattice Grassmann field with a quartic interaction \cite{Bugrij,BIS}, which disappears if the parameters 
 satisfy the condition
 \be\label{ffcondition}
 a_4=a_{12}a_{34}-a_{13}a_{24}+a_{14}a_{23}.
 \eb
 The model (\ref{gbweight}) can be mapped to the eight-vertex model in an 
 external field \cite{baxterbook}. The analog of the free-fermion condition (\ref{ffcondition}) in
 the vertex picture was obtained earlier in \cite{GreenHurst}.

 Below we study  the model defined by (\ref{gbweight})--(\ref{ffcondition}).
  It appeared in several equivalent formulations in papers by different authors.
 Its partition function can be evaluated by a variety of methods
 for both infinite and finite lattice \cite{BS2,Bugrij,FanWu,GreenHurst,Khachatryan}.
 Elliptic parametrization of the Boltzmann
 weights and inversion relations for the partition function were established in
 \cite{BS1,BS3}. Grassmann integral representations for the corresponding transfer matrix
  and its eigenvectors have been found in \cite{OL}. Infinite-lattice correlations in a dual model were studied in \cite{smj}.
 The model (\ref{gbweight})--(\ref{ffcondition}) includes as special cases the Ising
 model on the square and triangular lattices,
  as well as $\mathbb{Z}_n$-symmetric Baxter-Bazhanov-Stroganov $\tau^{(2)}$-model with $n=2$.

 % To authors' knowledge,
% this model was first introduced by Green and Hurst in \cite{GreenHurst}  and then
% studied and rediscovered in different forms by many authors,

 The transfer matrix $V_{\varepsilon}[\sigma,\sigma']$  is a function of $2N$ spin variables
 $\sigma_{0},\ldots,\sigma_{N-1}$, $\sigma'_{0},\ldots,\sigma'_{N-1}$ given by the product of plaquette weights
 (\ref{gbweight}) over one lattice row (Fig.~1):
 \be\label{tm}
 V_{\varepsilon}[\sigma,\sigma']=\prod_{j=0}^{N-1}W(\sigma_j,\sigma_j',\sigma_{j+1}',\sigma_{j+1}).
 \eb
 The subscript $\varepsilon=\pm1$ corresponds to periodic and antiperiodic boundary conditions on spin variables in the horizontal direction, i.e. $\sigma_N=\varepsilon\sigma_0$ and $\sigma_N'=\varepsilon\sigma_0'$.

  \begin{figure}[!h]
 \begin{center}
 \resizebox{7cm}{!}{
 \includegraphics{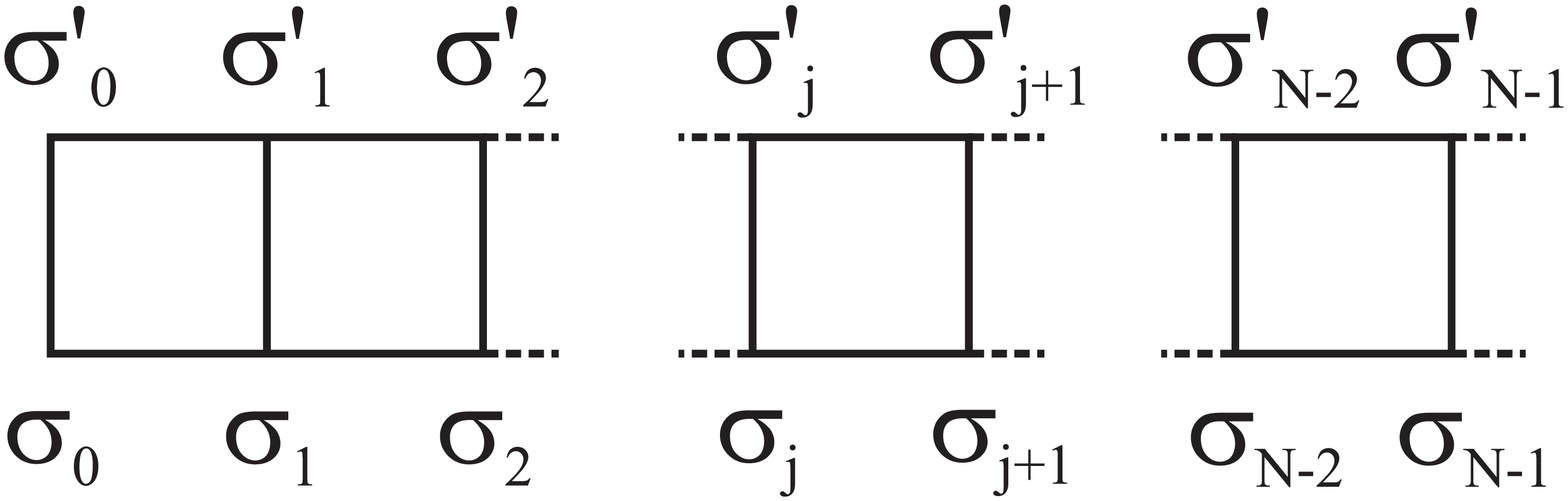}} \\
 Fig. 1
 \end{center}
 \end{figure}

 The transfer matrix acts on the $2^N$-dimensional space of
 maps $f:\left(\mathbb{Z}_2\right)^{\times N}\rightarrow \mathbb{C}$ by
 \ben
 (V_{\varepsilon}f)[\sigma]=\sum_{[\sigma']}V_{\varepsilon}[\sigma,\sigma']f[\sigma'].
 \ebn
 It commutes with the operators of translation and spin reflection defined by
 \begin{eqnarray*}
 (T_{\varepsilon}f)(\sigma_0,\ldots,\sigma_{N-1})=f(\sigma_1,\ldots,\sigma_{N-1},\varepsilon\sigma_0),
 \\ (Uf)[\sigma]=f[-\sigma].
 \end{eqnarray*}

 Partition function of the model can be written as $Z=\mathrm{Tr}\left(V_{\varepsilon}^MU^{\frac{1-\varepsilon'}{2}}\right)$, where $\varepsilon'=\pm1$
 corresponds to periodic and antiperiodic boundary conditions in the vertical direction.
 To write spin correlation functions, one should introduce spin operators $\{s_j\}$ with
 $j=0,\ldots,N-1$
 defined by
 $(s_jf)[\sigma]=\sigma_jf[\sigma]$. Then for $j_1\leq j_2\leq\ldots\leq j_n $ we have
 \ben
 \langle\sigma_{j_1,k_1}\ldots\sigma_{j_n,k_n}\rangle=Z^{-1}\mathrm{Tr}\left(
 s_{j_1,k_1}\ldots s_{j_n,k_n}V_{\varepsilon}^MU^{\frac{1-\varepsilon'}{2}}\right),
 \ebn
 where
 \ben
 s_{j,k}=V_{\varepsilon}^j s_k V_{\varepsilon}^{-j}=
 V_{\varepsilon}^j T_{\varepsilon}^k s_0 T_{\varepsilon}^{-k} V_{\varepsilon}^{-j}.
 \ebn
 The computation of correlation functions therefore reduces to
 finding matrix elements (form factors) of the spin operator $s_0$ in the common basis of eigenstates
 of $V_{\varepsilon}$, $T_{\varepsilon}$ and $U$. The present work is devoted to the solution
 of this problem.

  Multiplication of the Boltzmann weight (\ref{gbweight}) by $e^{K(\sigma_1\sigma_2-\sigma_3\sigma_4)}$
 with any $K$ does not change the transfer matrix. Thus, up to overall factor, $V_{\varepsilon}$ nontrivially depends on
 five parameters. The parameter set may be thought of as a five-dimensional projective
 space with homogeneous coordinates
  \begin{eqnarray}
 \label{kappa}\kappa=(a_{12}+a_{34})(a_{13}+a_{24})+(a_{14}+a_{23})(a_4+1),\\
 \label{lambda}\lambda=(a_{14}-a_{23})(a_{4}-1)-(a_{12}-a_{34})(a_{13}-a_{24}),\\
 \label{mu}\mu=(a_4+1)^2-(a_{12}+a_{34})^2-(a_{13}-a_{24})^2+(a_{14}-a_{23})^2,\\
 \label{xi}\varrho=4\left(a_{14}a_{23}-a_{13}a_{24}\right),\\
 \label{tau} \tau=(a_4+1)^2+(a_{12}+a_{34})^2+(a_{13}+a_{24})^2+(a_{14}+a_{23})^2,\\
 \label{upsilon} \upsilon=(a_{12}+a_{34})(a_{13}+a_{24})-(a_{14}+a_{23})(a_4+1).
 \end{eqnarray}
 It will be shown below that this set is foliated by two-dimensional
 surfaces such that the transfer matrices corresponding to different surface points
 are mutually commuting and diagonalized in the same basis. Different surfaces are labeled by the triples $\left(\frac{\lambda}{\kappa},
 \frac{\mu}{\kappa},\frac{\varrho}{\kappa}\right)$.
 Each of them contains a curve representing a set of transfer matrices
 of the Ising model on the triangular lattice, and this curve contains a point corresponding to
 the transfer matrix of an anisotropic Ising model on the square lattice.

 As a consequence, finite-lattice form factors of the free-fermion
 model (\ref{gbweight})--(\ref{ffcondition}) can be obtained  by a suitable parametrization from the Ising ones,
 for which the corresponding formulas were conjectured in \cite{BL1,BL2} and proved in \cite{Iorgov1,Iorgov2}.
 A more elegant proof based on earlier results of \cite{Hystad,Palmer_Hystad}
  and Frobenius determinant formula
 for elliptic Cauchy matrices was recently found in \cite{IL}.

 It is well-known that the six- and eight-vertex model transfer matrices commute with
the hamiltonians of the quantum XXZ and XYZ spin-1/2 chains, respectively. Similarly,
the dia\-go\-nal-to-diagonal transfer matrix of the Ising model on a square lattice commutes with the hamiltonian
of the quantum Ising chain  \cite{Onsager}. Local hamiltonians
are also known for the Ising model on the triangular and hexagonal lattices \cite{Stephen}.

We will show that
to every above-described two-parameter family of commuting transfer matrices
of the general free-fermion model
one can associate a local spin chain hamiltonian depending on three parameters
 $\frac{\lambda}{\kappa}$,
 $\frac{\mu}{\kappa}$, $\frac{\varrho}{\kappa}$.
It turns out to be related to the hamiltonian of
the quantum XY chain in a transverse field by a similarity tranformation. Simple considerations
 then allow us to rederive spin form factors of the finite-length
XY chain,  recently obtained in \cite{IorgovXY} by the method of separation of variables.

\section{Kaufman's approach and Grassmann integrals\label{tmdiag}}
 Let us define two operators $V_{a,p}$ by the relations
 \begin{eqnarray*}
 V_{\pm}=\frac{1\pm U}{2}V_a+\frac{1\mp U}{2}V_p.
 \end{eqnarray*}
 Since $V_{\pm}$ and $V_{a,p}$ commute with $U$, the set of eigenstates of $V_+$ ($V_-$)
 consists of even (odd) under spin reflection eigenvectors of $V_a$ and odd (resp. even)
 eigenvectors of $V_p$.

 It was observed in \cite{OL} that $V_{a,p}$ can be naturally represented as $2N$-fold Grassmann integrals.
 To describe this result in more detail, introduce
 auxiliary Grassmann variables $\psi_0,\ldots,\psi_{N-1}$, $\dot{\psi}_0,\ldots\dot{\psi}_{N-1}$.
 Our convention will be that ``dotted'' variables commute with the usual ones and anticommute between
 themselves. We will also need to consider discrete Fourier transforms of variables of both types
 \ben
 \psi_{\theta}=\frac{1}{\sqrt{N}}\sum_{j=0}^{N-1}e^{-ij\theta}\psi_j,
 \ebn
 using two sets of quasimomenta: $\boldsymbol{\theta}_{a}=\left\{\frac{\pi}{N},\frac{3\pi}{N},\ldots,2\pi-\frac{\pi}{N}\right\}$
  and  $\boldsymbol{\theta}_{p}=\left\{0,\frac{2\pi}{N},\ldots,2\pi-\frac{2\pi}{N}\right\}$.
 In this notation, one has \cite{OL}
 \be\label{tmgr}
 V_{\nu}[\sigma,\sigma']=\zeta_{\nu}\int\mathcal{D}\psi\,\mathcal{D}\dot{\psi}\;
 e^{\,S_{\nu}[\psi,\dot{\psi}]}\prod_{j=0}^{N-1}e^{\sigma_j\psi_j}
 \prod_{j=0}^{N-1}e^{\sigma'_j\dot{\psi}_j},\qquad \nu=a,p,
 \eb
 where $\zeta_{\nu}=a_0^N\left[\prod_{\theta\in\boldsymbol{\theta}_{\nu}}\chi_{\theta}\right]^{\frac12}$ and
 \begin{eqnarray}
 \label{actions}
 S_{\nu}[\psi,\dot{\psi}]=\frac12 \sum_{\theta\in\boldsymbol{\theta}_{\nu}}
 \left(\begin{array}{cc} \psi_{-\theta} & \dot{\psi}_{-\theta}\end{array}\right)
 \left(\begin{array}{cc} G_{11}(\theta) & G_{12}(\theta) \\ G_{21}(\theta) & G_{22}(\theta)
 \end{array}\right)\left(\begin{array}{c}\psi_{\theta}  \\ \dot{\psi}_{\theta}\end{array}\right),\\
 \label{chi}
 \fl \chi_{\theta}=\left[a_{12}+a_{34}+(a_{13}+a_{24})\cos\theta\right]^2+
 \left[(a_{13}-a_{24})^2+4a_{14}a_{23}\right]\sin^2\theta,\\
 \label{g11}
 \fl\chi_{\theta}G_{11}(\theta)=2i\sin\theta\left[a_{23}+a_{12}a_{24}+a_{13}a_{34}+a_{14}a_4-
 2\left(a_{14}a_{23}-a_{13}a_{24}\right)\cos\theta\right],\\
 \label{g22}
 \fl\chi_{\theta}G_{22}(\theta)=2i\sin\theta\left[a_{14}+a_{12}a_{13}+a_{24}a_{34}+a_{23}a_4-
 2\left(a_{14}a_{23}-a_{13}a_{24}\right)\cos\theta\right],\\
 \nonumber
 \fl\chi_{\theta}G_{12}(\theta)=\chi_{\theta}G_{21}(-\theta)=
 \left[(a_{12}+a_{34})(a_4+1)-(a_{13}+a_{24})(a_{14}+a_{23})\right]\\
 \label{g12} +\left[(a_{13}+a_{24})(a_4+1)-(a_{12}+a_{34})(a_{14}+a_{23})\right]\cos \theta\\
 \nonumber +\left[(a_{24}-a_{13})(a_4-1)+(a_{12}-a_{34})(a_{14}-a_{23})\right] i\sin\theta.
 \end{eqnarray}

 The representation (\ref{tmgr}) was used in \cite{OL} to find
 the eigenvectors of $V_{a,p}$ explicitly in terms of spin variables. One
 of the reasons to proceed in this way was the difficulties with the well-known Kaufman's algebraic
 approach \cite{kaufman} designed for the two-dimensional Ising model. It can be expected
 on general grounds that the conjugation by $V_{a,p}$ induces linear transformations
 of the standard Clifford algebra generators. However, in contrast to the Ising case, it was not clear how one can determine
 the explicit form of these transformations.

 This can be done as follows. First look for a representation
 for the inverse $V_{\nu}^{-1}[\sigma,\sigma']$ in the form (\ref{tmgr}):
 \be\label{invtmgr}
 {V}^{-1}_{\nu}[\sigma,\sigma']=\tilde{\zeta}_{\nu}\int\mathcal{D}\varphi\,\mathcal{D}\dot{\varphi}\;
 e^{\,\tilde{S}_{\nu}[\varphi,\dot{\varphi}]}\prod_{j=0}^{N-1}e^{\sigma_j\varphi_j}
 \prod_{j=0}^{N-1}e^{\sigma'_j\dot{\varphi}_j},
 \eb
 where the overall coefficient $\tilde{\zeta}_{\nu}$ and quadratic action
 $\tilde{S}_{\nu}[\varphi,\dot{\varphi}]$ should be chosen so that
 \be\label{invdef}
 \sum_{[\sigma']}V_{\nu}[\sigma,\sigma']V_{\nu}^{-1}[\sigma',\sigma'']=\delta_{[\sigma],[\sigma'']}
 =\prod_{j=0}^{N-1}\frac{1+\sigma^{\,}_j\sigma''_j}{2}.
 \eb
 Using (\ref{tmgr}) and the ansatz (\ref{invtmgr}), the l.h.s. of the last relation
 can be written as
 \be\label{vinvv}
 \fl\zeta_{\nu}\tilde{\zeta}_{\nu}\sum_{[\sigma']}\int\mathcal{D}\psi\,\mathcal{D}\dot{\psi}\,
 \mathcal{D}\varphi\,\mathcal{D}\dot{\varphi}\;
  e^{\,S_{\nu}[\psi,\dot{\psi}]}\prod_{j=0}^{N-1}e^{\sigma_j\psi_j}
 \prod_{j=0}^{N-1}e^{\sigma'_j\dot{\psi}_j}\;
 e^{\,\tilde{S}_{\nu}[\varphi,\dot{\varphi}]}\prod_{j=0}^{N-1}e^{\sigma'_j\varphi_j}
 \prod_{j=0}^{N-1}e^{\sigma''_j\dot{\varphi}_j}.
 \eb

 Let us explain how this expression can be simplified, since below
 similar calculations will be done without going into the details:
 \begin{itemize}
 \item Only even part of $\prod_{j=0}^{N-1}e^{\sigma_j\psi_j}
 \prod_{j=0}^{N-1}e^{\sigma'_j\dot{\psi}_j}$ gives non-zero contribution to the integral.
 Since it commutes with any quadratic expression in Grassmann variables of both types,
 the factor $e^{\,\tilde{S}_{\nu}[\varphi,\dot{\varphi}]}$ can be put together with
 $e^{\,S_{\nu}[\psi,\dot{\psi}]}$.
 \item The summation over intermediate spins can then be easily done:
 \ben
 \sum_{[\sigma']}\biggl(\prod_{j=0}^{N-1}e^{\sigma'_j\dot{\psi}_j}
 \prod_{j=0}^{N-1}e^{\sigma'_j\varphi_j}\biggr)=2^{N}e^{\dot{\psi}\varphi},
 \ebn
 where we use shorthand notation $\dot{\psi}\varphi=\sum_{j=0}^{N-1}\dot{\psi}_j\varphi^{\,}_j$.
 This transforms (\ref{vinvv}) into
 \ben
 2^N\zeta_{\nu}\tilde{\zeta}_{\nu}
 \int\mathcal{D}\psi\,\mathcal{D}\dot{\psi}\,
 \mathcal{D}\varphi\,\mathcal{D}\dot{\varphi}\;
  e^{\,S_{\nu}[\psi,\dot{\psi}]+\tilde{S}_{\nu}[\varphi,\dot{\varphi}]}
  \prod_{j=0}^{N-1}e^{\sigma_j\psi_j}\;e^{\dot{\psi}\varphi}\prod_{j=0}^{N-1}e^{\sigma''_j\dot{\varphi}_j}.
 \ebn
 \item To pull $e^{\dot{\psi}\varphi}$ to the left through $F[\psi]=\prod_{j=0}^{N-1}e^{\sigma_j\psi_j}$,
 write the latter factor as a sum of its odd and even part, which gives $F[\psi]e^{\dot{\psi}\varphi}=
 e^{\dot{\psi}\varphi}F_{\mathrm{even}}[\psi]+e^{-\dot{\psi}\varphi}F_{\mathrm{odd}}[\psi]$.
 In the integral corresponding to the first term, make the change of variables $\psi\rightarrow-\psi$,
 $\dot{\psi}\rightarrow-\dot{\psi}$. Since $S[\psi,\dot{\psi}]$ is even, this allows to rewrite
 the previous expression as
  \ben
 2^N\zeta_{\nu}\tilde{\zeta}_{\nu}
 \int\mathcal{D}\psi\,\mathcal{D}\dot{\psi}\,
 \mathcal{D}\varphi\,\mathcal{D}\dot{\varphi}\;
  e^{\,S_{\nu}[\psi,\dot{\psi}]+\tilde{S}_{\nu}[\varphi,\dot{\varphi}]-\dot{\psi}\varphi}
  \prod_{j=0}^{N-1}e^{\sigma_j\psi_j}\prod_{j=0}^{N-1}e^{\sigma''_j\dot{\varphi}_j}.
 \ebn
 \end{itemize}
 Thus, in order to satisfy (\ref{invdef}), it is sufficient to choose $\tilde{\zeta}_{\nu}$
 and $\tilde{S}_{\nu}[\varphi,\dot{\varphi}]$ so that
   \ben
 2^{2N}\zeta_{\nu}\tilde{\zeta}_{\nu}
 \int\mathcal{D}\dot{\psi}\, \mathcal{D}\varphi\;
  e^{\,S_{\nu}[\psi,\dot{\psi}]+\tilde{S}_{\nu}[\varphi,\dot{\varphi}]-\dot{\psi}\varphi}
  =e^{\psi\dot{\varphi}}.
 \ebn
 In view of (\ref{actions}), it is natural to try to satisfy this relation with $\tilde{S}_{\nu}[\varphi,\dot{\varphi}]$ diagonalized
 by discrete Fourier transform. This indeed works and one finds that
  \begin{eqnarray*}
 \tilde{S}_{\nu}[\varphi,\dot{\varphi}]=\frac12 \sum_{\theta\in\boldsymbol{\theta}_{\nu}}
 \left(\begin{array}{cc} \dot{\varphi}_{-\theta} & {\varphi}_{-\theta}\end{array}\right)
 G^{-1}(\theta)\left(\begin{array}{c}\dot{\varphi}_{\theta}  \\ {\varphi}_{\theta}\end{array}\right),\\
 \tilde{\zeta}_{\nu}^{-1}=2^{2N}\zeta_{\nu}\biggl[\,\prod_{\theta\in\boldsymbol{\theta}_{\nu}}
 \frac{G_{12}(\theta)G_{21}(\theta)}{\mathrm{det}\,G(\theta)}\biggr]^{\frac12},
 \end{eqnarray*}
 where the $2\times2$ matrix $G(\theta)$ is defined by (\ref{chi})--(\ref{g12}).

 The next step is to introduce the Clifford algebra generators $\{p_j\}$, $\{q_j\}$  ($j=0,\ldots,N-1$),
 \begin{eqnarray*}
 \left(\begin{array}{c} p_j \\ q_j \end{array}\right)[\sigma,\sigma']=2^{-N}
 \prod_{k=0}^{j-1}(1-\sigma^{\,}_k\sigma'_k)
 \prod_{k=j+1}^{N-1}(1+\sigma^{\,}_k\sigma'_k)
 \left(\begin{array}{c}\sigma^{\,}_j+\sigma'_j \\ i(\sigma'_j-\sigma^{\,}_j)\end{array}\right),
 \end{eqnarray*}
 which satisfy standard anticommutation relations
 $ \left\{p_j,p_k\right\}= \left\{q_j,q_k\right\}=2\delta_{jk}$, $\left\{p_j,q_k\right\}=0$.
 It is easy to verify that these operators can be represented in a form similar to Grassmann
 integral representation (\ref{tmgr}):
 \be\label{pqgr}
 \fl\qquad\qquad \left(\begin{array}{c} p_j \\ q_j \end{array}\right)[\sigma,\sigma']=2^{-N}
 \int\mathcal{D}\eta\,\mathcal{D}\dot{\eta}\;
 e^{\,\eta\dot{\eta}}\left(\begin{array}{c} \eta_j+\dot{\eta}_j \\
 i(\eta_j-\dot{\eta}_j) \end{array}\right)
 \prod_{k=0}^{N-1}e^{\sigma_k\eta_k}
 \prod_{k=0}^{N-1}e^{\sigma'_k\dot{\eta}_k}.
 \eb
 We especially note that Fourier transforms $p_{\theta}$, $q_{\theta}$ are obtained
 by simple replacement of the subscript $j$ by $\theta$ in the r.h.s. of (\ref{pqgr}).

 It then becomes possible to compute the result
 of conjugation of $\{p_j\}$, $\{q_j\}$ by $V_{\nu}$
 using the representations (\ref{tmgr}), (\ref{invtmgr}) and (\ref{pqgr}). After summation
 over intermediate spin variables one obtains
 \begin{eqnarray*}
 V^{\,}_{\nu}\left(\begin{array}{c} p_{\theta} \\ q_{\theta} \end{array}\right)V_{\nu}^{-1}[\sigma,\sigma']
 =2^{N}\zeta_{\nu}\tilde{\zeta}_{\nu}
 \int\mathcal{D}\psi\,\mathcal{D}\dot{\psi}\,\mathcal{D}\eta\,\mathcal{D}\dot{\eta}\,
 \mathcal{D}\varphi\,\mathcal{D}\dot{\varphi}\\
 e^{\,S_{\nu}[\psi,\dot{\psi}]+
 \tilde{S}_{\nu}[\varphi,\dot{\varphi}]-\dot{\psi}\eta+\eta\dot{\eta}-\dot{\eta}\varphi}\left(\begin{array}{c} \eta_{\theta}+\dot{\eta}_{\theta} \\
  i(\eta_{\theta}-\dot{\eta}_{\theta})\end{array}\right)
  \prod_{j=0}^{N-1}e^{\sigma_j\psi_j}
 \prod_{j=0}^{N-1}e^{\sigma'_j\dot{\varphi}_j}.
 \end{eqnarray*}
 The integration over $\dot{\psi}$, $\eta$, $\dot{\eta}$, $\varphi$ can be
 performed in the  Fourier basis. Comparing the resulting integral over $\psi$ and $\dot{\varphi}$
 with (\ref{pqgr}), we find that the induced
 rotation is explicitly given by
 \begin{eqnarray}\label{pqrotation1}
 V^{\,}_{\nu}\left(\begin{array}{c} p_{\theta} \\ q_{\theta} \end{array}\right)V_{\nu}^{-1}
 =\Lambda(V_{\nu})
 \left(\begin{array}{c} p_{\theta} \\ q_{\theta} \end{array}\right),\\
  \label{pqrotation2}\Lambda(V_{\nu})=\frac{1}{2\chi_{\theta}G_{12}(\theta)}\left(\begin{array}{cc} \alpha_{\theta} & i\beta_{-\theta} \\
 -i\beta_{\theta} &  \alpha_{-\theta} \end{array}\right),
 \end{eqnarray}
 where
 \begin{eqnarray}
 \label{alphabeta}
 \cases{
 \alpha_{\theta}=\chi_{\theta}\left[1-\mathrm{det}\,G(\theta)+G_{11}(\theta)-G_{22}(\theta)\right],\\
 \beta_{\theta}=\chi_{\theta}\left[1+\mathrm{det}\,G(\theta)+G_{11}(\theta)+G_{22}(\theta)\right].
 }
 \end{eqnarray}
 One can now follow Kaufman's method and find the eigenstates of $V_{a,p}$ by diagonalization of
 the two-dimensional rotations (\ref{pqrotation2}).

 It is straightforward to check using (\ref{chi})--(\ref{g12}) that $\alpha_{\theta}$, $\beta_{\theta}$
  can be written in terms of the parameters (\ref{kappa})--(\ref{upsilon}) as follows:
 \begin{eqnarray}
 \label{detg}
 \alpha_{\theta}=\tau+2\upsilon\cos\theta+2i\lambda\sin\theta,\qquad
 \beta_{\theta}=-\varrho e^{2i\theta}+2\kappa e^{i\theta}-\mu.
 \end{eqnarray}
 and that $\alpha_{\theta}\alpha_{-\theta}-\beta_{\theta}\beta_{-\theta}=
 4\chi_{\theta}^2G_{12}(\theta)G_{21}(\theta)$. In fact the relations (\ref{detg})
 are the origin of the parametrization (\ref{kappa})--(\ref{upsilon}). Let us
 further introduce
 three parameters $\mathcal{K}_0$, $\mathcal{K}_x$, $\mathcal{K}_y$ by
 \begin{eqnarray*}
 \fl\qquad\frac{\lambda}{\kappa}=\frac{\sinh2\Kr_0}{\cosh2\Kr_x\sinh2\Kr_y},\quad
 \frac{\mu}{\kappa}=\frac{\cosh2\Kr_y+\cosh2\Kr_0}{\cosh2\Kr_x\sinh2\Kr_y},\quad
 \frac{\varrho}{\kappa}=\frac{\cosh2\Kr_y-\cosh2\Kr_0}{\cosh2\Kr_x\sinh2\Kr_y}.
 \end{eqnarray*}
 It will be shown in the next section
 that the eigenvectors of $V_{\varepsilon}$ coincide with the eigenvectors of a
 non-symmetrized transfer matrix of the anisotropic Ising model on the square lattice:
 \be\label{nstm}
 \fl\qquad\exp\biggl\{\frac{\Kr_y-\Kr_0}{2}\sum_{j=0}^{N-1}s_js_{j+1}\biggr\}
 \exp\biggl\{\Kr_x^*\sum_{j=0}^{N-1}C_j\biggr\}
 \exp\biggl\{\frac{\Kr_y+\Kr_0}{2}\sum_{j=0}^{N-1}s_js_{j+1}\biggr\},
 \eb
 where $\tanh \Kr_x^*=e^{-2\Kr_x}$ and the operators $\{C_j\}$ are defined by
  \ben
 (C_jf)(\ldots,\sigma_{j-1},\sigma_j,\sigma_{j+1},\ldots)=f(\ldots,\sigma_{j-1},-\sigma_j,\sigma_{j+1},\ldots).
 \ebn
 We assume for concreteness that $\mathcal{K}_0$, $\mathcal{K}_x$, $\mathcal{K}_y$ are real and positive
 and  $\Kr_x^*<\Kr_y$. This mimics the ferromagnetic region of Ising parameters.

 Define the functions
 \ben
 b^{\pm}_{\theta}=\frac{\sqrt{(\alpha_{\theta}-\alpha_{-\theta})^2+4\beta_{\theta}\beta_{-\theta}}\pm
 (\alpha_{\theta}-\alpha_{-\theta})}{2e^{-i\theta}\beta_{\theta}},
 \ebn
 satisfying $b^{\pm}_{\theta}b^{\pm}_{-\theta}=1$. The square roots here and below
 are fixed by the requirement of positivity of the real part.
 Determining the eigenvectors of the
  induced rotations (\ref{pqrotation2}), introduce the creation-annihilation operators
 \begin{eqnarray}\label{caops}
 \cases{
 2\psi^{\dag}_{\theta}=\rho_{\theta}\left(
 e^{-i\theta}\sqrt{b^+_{\theta}}\,p_{-\theta}-i\sqrt{b^+_{-\theta}}\,q_{-\theta}\right),\vspace{0.1cm}\\
 2\psi_{\theta}=\rho_{\theta}\left(e^{i\theta}\sqrt{b^-_{-\theta}}\,p_{\theta}\,+
 \,i\sqrt{b^-_{\theta}}\,q_{\theta}\right),
 }
 \end{eqnarray}
 where
 \ben
 \rho_{\theta}=\rho_{-\theta}=\sqrt2\left[\sqrt{b^+_{\theta}b^-_{-\theta}}+
 \sqrt{b^-_{\theta}b^+_{-\theta}}\right]^{-\frac12}.
 \ebn

 These operators satisfy canonical anticommutation relations
 $ \{\psi^{\dag}_{\theta},\psi^{\dag}_{\theta'}\}=\{\psi_{\theta},\psi_{\theta'}\}=0$,
 $ \{\psi^{\dag}_{\theta},\psi_{\theta'}\}=\delta_{\theta,\theta'}$.
% and  the inversion formulas are given by
% \begin{eqnarray}\label{inversionfs}
% p_{\theta}=\rho_{\theta}e^{-i\theta}\left(\sqrt{b^-_{\theta}}\,\psi^{\dag}_{-\theta}+
% \sqrt{b^+_{\theta}}\,\psi_{\theta}\right),\qquad
% q_{\theta}=i\rho_{\theta}\left(\sqrt{b^-_{-\theta}}\,\psi^{\dag}_{-\theta}-
% \sqrt{b^+_{-\theta}}\,\psi_{\theta}\right).
% \end{eqnarray}
  They transform diagonally under conjugation by $V_{\nu}$:
 \ben
 V_{\nu} \left(\begin{array}{c}
 \psi^{\dag}_{\theta}  \\
 \psi_{\theta}
 \end{array}\right)V_{\nu}^{-1}=\left(\begin{array}{cc}
 e^{-\mathcal{E}_{\theta}} & 0 \\
 0 & e^{\mathcal{E}_{\theta}}
 \end{array}\right)
 \left(\begin{array}{c}
 \psi^{\dag}_{\theta}  \\
 \psi_{\theta}
 \end{array}\right),\qquad \theta\in\boldsymbol{\theta}_{\nu},
 \ebn
 where
 \be\label{scurve1}
 \mathcal{E}_{\theta}=\ln\frac{\sqrt{(\alpha_{\theta}-\alpha_{-\theta})^2+4\beta_{\theta}\beta_{-\theta}}
 +\alpha_{\theta}+\alpha_{-\theta}}{4\chi_{\theta}G_{12}(\theta)}.
 \eb
 The matrices $V_{\nu}$ can therefore be written as
 \be\label{tmppb}
 V_{\nu}=2^Na_0^N\biggl[\,\prod_{\theta\in\boldsymbol{\theta}_{\nu}}\chi_{\theta}G_{12}(\theta)\biggr]^{\frac12}
 \exp\biggl\{-\sum_{\theta\in\boldsymbol{\theta}_{\nu}}
 \mathcal{E}_{\theta}\left(\psi^{\dag}_{\theta}\psi^{\,}_{\theta}-
 \frac12)\right)\biggr\}.
 \eb
  where the overall scalar multiple can be
 fixed e.g. by the identification of eigenvalues with formulas in \cite{OL}.
 In contrast to the Ising case, $\mathcal{E}_{\theta}\neq\mathcal{E}_{-\theta}$ and hence
 the spectrum is (generically) nondegenerate. In particular, the eigenstates of $V_{\varepsilon}$
 will automatically diagonalize the translation operator.

 The left and right eigenvectors of $V_{\nu}$ are multiparticle Fock states
 \begin{eqnarray}
 \label{fockstates}
 \fl\qquad_{\nu}\langle \theta_1,\ldots,\theta_k|= _{\nu}\langle vac|\psi_{\theta_1}\ldots\psi_{\theta_k},
 \qquad
 |\theta_1,\ldots,\theta_k\rangle_{\nu}=\psi^{\dag}_{\theta_1}\ldots\psi^{\dag}_{\theta_k}|vac\rangle_{\nu},
 \end{eqnarray}
 where $\theta_1,\ldots,\theta_k\in\boldsymbol{\theta}_{\nu}$. The corresponding eigenvalue
  is equal to $\exp\left\{\frac12\sum_{\theta\in\boldsymbol{\theta}_{\nu}}\mathcal{E}_{\theta}-\sum_{i=1}^k\mathcal{E}_{\theta_i}\right\}$. Vacuum vectors $|vac\rangle_{\nu}$
 and $_{\nu}\langle vac|$ are annihilated by all $\psi_{\theta}$ (resp. $\psi^{\dag}_{\theta}$) with $\theta\in\boldsymbol{\theta}_{\nu}$
 and are normalized as $_{\nu}\langle vac|vac\rangle_{\nu}=1$. Since $V$ is not symmetric, $_{\nu}\langle vac|$ is not necessarily hermitian conjugate of $|vac\rangle_{\nu}$.

 Not all of the states
 (\ref{fockstates}) are eigenvectors of the full transfer matrix (and the
 associated quantum spin chain hamiltonians below).
 For periodic and
 antiperiodic boundary conditions on spin variables the number of particles in these states
 should be even and odd, respectively, and one has
  \begin{eqnarray}
  \cases{
  T_+|\theta_1,\ldots,\theta_{2k}\rangle_{a,p}\quad=
 e^{-i\sum_{i=1}^{2k}\theta_i}\;|\theta_1,\ldots,\theta_{2k}\rangle_{a,p},\\
 T_-|\theta_1,\ldots,\theta_{2k+1}\rangle_{a,p}=
 e^{-i\sum_{i=1}^{2k+1}\theta_i}|\theta_1,\ldots,\theta_{2k+1}\rangle_{a,p}.
 }
 \end{eqnarray}

  We especially note that the transfer matrix eigenvectors and form factors depend only on
  $\Kr_x$, $\Kr_y$, $\Kr_0$ and are independent of $\frac{\tau}{\kappa}$, $\frac{\upsilon}{\kappa}$.
  The latter two variables appear only in the eigenvalues and can be thought of as spectral parameters.
  \section{Form factors}
  Consider instead of $V_{\varepsilon}$ the conjugated transfer matrix
  $V_{\varepsilon}'=SV_{\varepsilon} S^{-1}$ with
  $S=\exp\,\Bigl\{\frac{\Kr_0}{2} \sum_{j=0}^{N-1}s_j s_{j+1}\Bigr\}$.
  The matrix of induced rotation in (\ref{pqrotation1}) then becomes
  $\Lambda(V'_{\nu})=\Lambda^{-1}(S)\Lambda(V_{\nu})\Lambda(S)$ with
  \ben
  \Lambda(S)=\left(\begin{array}{cc}
  \cosh\Kr_0 & i\sinh\Kr_0\,e^{-i\theta} \\ -i\sinh\Kr_0\,e^{i\theta} & \cosh\Kr_0
  \end{array}\right),
  \ebn
  and it can be straightforwardly checked that $\Lambda(V'_{\nu})=\Lambda(V_{\nu})\Bigl|_{\Kr_0=0}\Bigr.$.
  Therefore the orthonormal system of left and right eigenvectors of $V_{\nu}$ can be
  chosen as follows:
  \ben
  \fl |\theta_1,\ldots,\theta_k\rangle_{\nu}^{S}=S^{-1}\Bigl(|\theta_1,\ldots,\theta_k\rangle_{\nu}\Bigr)_{\Kr_0=0},
  \qquad
  {}_{\nu}^{S}\langle\theta_1,\ldots,\theta_k|=
  \Bigl({}_{\nu}\langle\theta_1,\ldots,\theta_k|\Bigr)_{\Kr_0=0}S.
  \ebn
  These vectors are of course proportional to those in (\ref{fockstates}).

  Since $S$ commutes with the operators $\{s_j\}$, nontrivial spin form factors coincide
  with those computed for $\Kr_0=0$:
  \begin{eqnarray}
  \nonumber\fl\mathcal{F}^{(l)}_{m,n}(\boldsymbol{\theta},\boldsymbol{\theta}')=&
  {}_{a}^{S}\langle\theta_1,\ldots,\theta_m|s_l|\theta'_1,\ldots,\theta'_n\rangle_{p}^S=\left[
  \label{ffwis}{}_{p}^{S}\langle\theta'_1,\ldots,\theta'_n|s_l|\theta_1,\ldots,\theta_m\rangle_{a}^S\right]^*=\\
  =&\left[{}_{a}\langle\theta_1,\ldots,\theta_m|s_l|\theta'_1,\ldots,\theta'_n\rangle_{p}\right]_{\Kr_0=0},
  \end{eqnarray}
  where $m$, $n$ are simultaneously even or odd.
  On the other hand, setting $\Kr_0=0$ leads to important simplifications
  in the formulas of the previous section. It implies in particular that $\lambda=0$, $\alpha_{\theta}=\alpha_{-\theta}$,
  $b^+_{\theta}=b^{-}_{\theta}\stackrel{def}{ = }b_{\theta}$ and $\rho_{\theta}=1$. Moreover,
  \be\label{btheta}
 b_{\theta}=\sqrt{\frac{\left(1-\tanh\Kr_x^*\coth\Kr_y e^{i\theta}\right)\left(1-\tanh\Kr_x^*\tanh\Kr_y e^{-i\theta}\right)}{
 \left(1-\tanh\Kr_x^*\tanh\Kr_y e^{i\theta}\right)\left(1-\tanh\Kr_x^*\coth\Kr_y e^{-i\theta}\right)}},
 \eb
 and thus the Fock states (\ref{fockstates}) with $\Kr_0=0$ coincide with the eigenvectors of
 the symmetric transfer matrix of the anisotropic Ising model on the square lattice, cf. \cite{IL}.
 Let us therefore denote $|\theta_1,\ldots,\theta_k\rangle_{\nu}^{\mathrm{Ising}}=\Bigl(|\theta_1,\ldots,\theta_k\rangle_{\nu}\Bigr)_{\Kr_0=0}$.
 In the symmetric Ising case, the left and right eigenvectors are related by hermitian conjugation, i.e. ${}^{\mathrm{Ising}}_{
 \quad\nu}
 \langle\theta_k,\ldots,\theta_1|= \left(|\theta_1,\ldots,\theta_k\rangle_{\nu}^{\mathrm{Ising}}\right)^{\dag}$.

  To write the corresponding form factors,
  introduce two functions $\gamma_{\theta}$, $\nu_{\theta}$ and two parameters $\xi$, $\xi_T$
  defined by
  \ben
  \cosh\gamma_{\theta}=\cosh2\Kr_x^*\cosh2\Kr_y-\sinh2\Kr_x^*\sinh2\Kr_y\cos\theta,\qquad
  \gamma_{\theta}>0,
  \ebn
  \ben
  \nu_{\theta}=\ln\frac{\prod_{\theta'\in\boldsymbol{\theta}_a}
  \sinh\frac{\gamma_{\theta}+\gamma_{\theta'}}{2}}{\prod_{\theta'\in\boldsymbol{\theta}_p}
  \sinh\frac{\gamma_{\theta}+\gamma_{\theta'}}{2}},\qquad \xi=\left[1-(\sinh2\Kr_x\sinh2\Kr_y)^{-2}\right]^{\frac14},
  \ebn
  \ben
   \xi_T=\prod_{\theta\in\boldsymbol{\theta}_p} e^{\nu_{\theta}/4}
 \prod_{\theta\in\boldsymbol{\theta}_a} e^{-\nu_{\theta}/4}=
 \left[\frac{\prod_{\theta\in\boldsymbol{\theta}_p}\prod_{\theta'\in\boldsymbol{\theta}_a}
 \sinh^2\frac{\gamma_{\theta}+\gamma_{\theta'}}{2}}{\prod_{\theta,\theta'\in\boldsymbol{\theta}_p}
 \sinh\frac{\gamma_{\theta}+\gamma_{\theta'}}{2}\prod_{\theta,\theta'\in\boldsymbol{\theta}_a}
 \sinh\frac{\gamma_{\theta}+\gamma_{\theta'}}{2}}\right]^{\frac14}.
 \ebn
 Then, by (\ref{ffwis}) and e.g. Theorem~6.1 in \cite{IL}:
   \begin{eqnarray}
 \label{isingff}
 \fl\mathcal{F}_{m,n}^{(l)}(\boldsymbol{\theta},\boldsymbol{\theta}')=i^{2mn-\frac{ m+n}{2}}\sqrt{\xi\xi_T}
 \prod_{j=1}^m
 \frac{e^{-i(l-\frac12)\theta_j+\nu_{\theta_j}/2}}{\sqrt{N\sinh\gamma_{\theta_j}}}
 \prod_{j=1}^n
 \frac{e^{i(l-\frac12)\theta'_j-\nu_{\theta'_j}/2}}{\sqrt{N\sinh\gamma_{\theta'_j}}}\;\times\\
 \fl\nonumber\times\left(\frac{\sinh2\Kr_y}{\sinh2\Kr_x}\right)^{\frac{(m-n)^2}{4}}\!\!\!\!\!\!\prod_{1\leq i<j\leq m}\frac{\sin\frac{\theta_i-\theta_j}{2}}{\sinh\frac{\gamma_{\theta_i}+\gamma_{\theta_j}}{2}}
  \prod_{1\leq i<j\leq n}\frac{\sin\frac{\theta'_i-\theta'_j}{2}}{\sinh\frac{\gamma_{\theta'_i}+\gamma_{\theta'_j}}{2}}
  \prod_{1\leq i\leq m,1\leq j\leq n}\frac{\sinh\frac{\gamma_{\theta_i}+\gamma_{\theta'_j}}{2}
  }{\sin\frac{\theta_i-\theta'_j}{2}}.
 \end{eqnarray}
 As explained in the Introduction, this formula allows to compute (long-distance expansions of)
 an arbitrary spin correlation function in the statistical model (\ref{gbweight})--(\ref{ffcondition}).

 \section{Quantum spin chain hamiltonian}
 The matrix $V_{\nu}$ clearly commutes (cf (\ref{tmppb})) with the operator
 \ben
 H_{\nu}=-\frac{1}{2}\sum_{\theta\in\boldsymbol{\theta}_{\nu}}\left\{\frac{4\sqrt{\beta_{\theta}
 \beta_{-\theta}}}{\rho_{\theta}^2}\,\psi^{\dag}_{\theta}\psi^{\,}_{\theta}-
 \sqrt{(\alpha_{\theta}-\alpha_{-\theta})^2-4\beta_{\theta}\beta_{-\theta}}\right\},
 \ebn
 which has the following form in terms of $\{p_{\theta}\}$, $\{q_{\theta}\}$:
 \ben
 H_{\nu}=-\frac14\sum_{\theta\in\boldsymbol{\theta}_{\nu}}\Bigl\{
 \left(\alpha_{\theta}-\alpha_{-\theta}\right)\left(p_{-\theta}p_{\theta}-q_{-\theta}q_{\theta}\right)
 -4i\beta_{\theta}\,q_{-\theta}p_{\theta}\Bigr\}.
 \ebn
 Now using (\ref{alphabeta})
 and performing the inverse Fourier transform, one finds that
 \ben
 H_{\nu}=-\sum_{j=0}^{N-1}\Bigl\{\lambda\left(p_jp_{j+1}-q_{j}q_{j+1}\right)+i\mu\, q_j p_j-
 2i\kappa\, q_jp_{j+1}+i\varrho\,q_{j}p_{j+2}\Bigr\},
 \ebn
 where $p_{j+N}=p_j$, $q_{j+N}=q_j$ for $\nu=p$ and $p_{j+N}=-p_j$, $q_{j+N}=-q_j$ for $\nu=a$.

 Using the standard realization of $\{p_j\}$, $\{q_j\}$ in terms of Pauli matrices $\sigma^{x,y,z}$,
 \begin{eqnarray*}
 p_j=\underbrace{\sigma^x\otimes\ldots\otimes\sigma^x}_{j\;\mathrm{times}}\otimes
 \,\sigma^z\otimes\underbrace{\mathbf{1}\otimes
 \ldots\otimes\mathbf{1}}_{N-1-j\;\mathrm{ times}},\\
 q_j=\underbrace{\sigma^x\otimes\ldots\otimes\sigma^x}_{j\;\mathrm{times}}\otimes
 \,\sigma^y\otimes\underbrace{\mathbf{1}\otimes
 \ldots\otimes\mathbf{1}}_{N-1-j \;\mathrm{ times}},
 \end{eqnarray*}
  one finds the quantum spin chain hamiltonian, which
 commutes with the transfer matrix $V_{\varepsilon}$
 and the operator (\ref{nstm}):
 \begin{eqnarray*}
 H_{\varepsilon}=\frac{1+\varepsilon U}{2}\,H_a+\frac{1-\varepsilon U}{2}\,H_p
 =\\ = \sum_{j=0}^{N-1}\Bigl\{
 2\kappa \sigma^z_j\sigma^z_{j+1}+\mu\sigma^x_j-
 i\lambda\left(\sigma^y_{j}\sigma^z_{j+1}+\sigma^z_j\sigma^y_{j+1}\right)
 -\varrho\, \sigma^z_j\sigma^x_{j+1}\sigma^z_{j+2}\Bigr\},
 \end{eqnarray*}
 where the boundary conditions are $\sigma^x_{j+N}=\sigma^x_j$,
 $\sigma^y_{j+N}=\varepsilon\sigma^y_j$,  $\sigma^z_{j+N}=\varepsilon\sigma^z_j$. To put this
 hamiltonian in a more familiar form, conjugate it by the matrix $S'=V_x^{\frac12}V_y^{\frac12}S$ with
 \ben
 V_x=\exp\,\Bigl\{\Kr_x^*\sum_{j=0}^{N-1}\sigma^x_j\Bigr\},\qquad
 V_y=\exp\,\Bigl\{\Kr_y\sum_{j=0}^{N-1}\sigma^z_j\sigma^z_{j+1}\Bigr\},
 \ebn
 which gives
 \begin{eqnarray}
 \label{hxy}
 \fl\qquad H'_{\varepsilon}=S'H_{\varepsilon}S'^{-1}=\mathrm{const}\cdot\sum_{j=0}^{N-1}\Bigl\{
 e^{-2\Kr_x}\sigma^y_j\sigma^y_{j+1}+
 e^{2\Kr_x}\sigma^z_j\sigma^z_{j+1}+
 2\coth2\Kr_y\,\sigma^x_j\Bigr\}.
 \end{eqnarray}
 One recognizes here the hamiltonian of the XY chain in a transverse field. If we restrict
 ourselves to the symmetric Ising transfer matrix
 by setting $V_{\varepsilon}=V_y^{\frac12}V_xV_y^{\frac12}$, this reproduces the well-known observation
 of \cite{suzuki} that $H'_{\varepsilon}$ commutes with the second symmetric Ising transfer
 matrix $V_x^{\frac12}V_yV_x^{\frac12}$.

 The orthonormal eigenvectors
 of $H'_{\varepsilon}$ can therefore be chosen as follows:
 \begin{eqnarray}
 \label{rvxy}
 |\theta_1,\ldots,\theta_k\rangle_{\nu}^{\mathrm{XY}}=
 e^{\pm\frac{\gamma(\boldsymbol{\theta})}{2}}V_x^{\pm\frac12}
 V_y^{\pm\frac12}|\theta_1,\ldots,\theta_k\rangle_{\nu}^{\mathrm{Ising}},\\
 \label{lvxy}
 {}_{\;\;\nu}^{\mathrm{XY}}\langle \theta_1,\ldots,\theta_k|=
 {}_{\quad\nu}^{\mathrm{Ising}}\langle \theta_1,\ldots,\theta_k|
 V_y^{\pm\frac12}V_x^{\pm\frac12}e^{\pm\frac{\gamma(\boldsymbol{\theta})}{2}},
 \end{eqnarray}
 where
  \be\label{gammabold}
 e^{-\gamma(\boldsymbol{\theta})}=\exp\Bigl\{\frac12\sum_{\theta\in\boldsymbol{\theta}_{\nu}}\gamma_{\theta}
 -\sum_{i=1}^k\gamma_{\theta_i}\Bigr\}
 \eb
 is the corresponding eigenvalue of  $V_x^{\frac12}V_yV_x^{\frac12}$.
 Note that different signs in (\ref{rvxy}) and (\ref{lvxy}) give equivalent representations
 of the same vector, and we have
 the hermitian conjugation identities ${}_{\;\;\nu}^{\mathrm{XY}}\langle \theta_k,\ldots,\theta_1|=\left(
 |\theta_1,\ldots,\theta_k\rangle_{\nu}^{\mathrm{XY}}\right)^{\dag}$.

 The operators $\sigma^x_l=ip_lq_l$  commute with the $\Zb_2$-charge $U$
 and thus have non-zero form factors only between the states of the same type ($a$- or $p$-).
 Since they are  bilinear in fermions, their matrix elements are easily computable.
 The only nontrivial form factors of the XY chain
 (\ref{hxy}) are those of the operators  $\sigma^y_l$ and $\sigma^z_l$, which map the $a$-sector to
 $p$-sector and vice versa. Define
 \begin{eqnarray*}
 \mathcal{F}^{\mathrm{XY}}_{m,n}(\boldsymbol{\theta},
 \boldsymbol{\theta}'|\sigma^{r}_l)&=
 \,\;{}_{\;\;\;a}^{\mathrm{XY}}\langle \theta_1,\ldots,\theta_m|\sigma^{r}_l|
 \theta'_1,\ldots,\theta'_n\rangle_{p}^{\mathrm{XY}}=\\
 &=\left[\,
 {}_{\;\;\;p}^{\mathrm{XY}}\langle \theta'_1,\ldots,\theta'_n|\sigma^{r}_l|
 \theta_1,\ldots,\theta_m\rangle_{a}^{\mathrm{XY}}\right]^*,\qquad r=y,z.
 \end{eqnarray*}
 Then, using different representations for the eigenvectors (\ref{rvxy})--(\ref{lvxy}),
 one finds that
 \begin{eqnarray*}
 \fl\pm i\sinh\Kr_x^* \mathcal{F}^{\mathrm{XY}}_{m,n}(\boldsymbol{\theta},
 \boldsymbol{\theta}'|\sigma^{y}_l)+\cosh\Kr_x^*
 \mathcal{F}^{\mathrm{XY}}_{m,n}(\boldsymbol{\theta},
 \boldsymbol{\theta}'|\sigma^{z}_l)=e^{\pm\frac{\gamma(\boldsymbol{\theta})-\gamma(\boldsymbol{\theta}')}{2}}
 \times\\
 \fl\times\;
 {}_{\quad a}^{\mathrm{Ising}}\langle \theta_1,\ldots,\theta_m|V_y^{\pm\frac12}V_x^{\pm\frac12}
 (\pm i\sinh\Kr_x^*\sigma^y_l+\cosh\Kr_x^*\sigma^z_l)
 V_x^{\mp\frac12}V_y^{\mp\frac12}|\theta'_1,\ldots,\theta'_n\rangle_{p}^{\mathrm{Ising}}=\\
 \fl=e^{\pm\frac{\gamma(\boldsymbol{\theta})-\gamma(\boldsymbol{\theta}')}{2}}\;
 {}_{\quad a}^{\mathrm{Ising}}\langle \theta_1,\ldots,\theta_m|V_y^{\pm\frac12}
 \sigma^z_l
 V_y^{\mp\frac12}|\theta'_1,\ldots,\theta'_n\rangle_{p}^{\mathrm{Ising}}=
 e^{\pm\frac{\gamma(\boldsymbol{\theta})-\gamma(\boldsymbol{\theta}')}{2}}
 \mathcal{F}^{(l)}_{m,n}(\boldsymbol{\theta},
 \boldsymbol{\theta}').
 \end{eqnarray*}
 From these relations one obtains the final finite-length XY form factor formulas:
 \begin{eqnarray}
 \label{xyff1}
 \mathcal{F}^{\mathrm{XY}}_{m,n}(\boldsymbol{\theta},
 \boldsymbol{\theta}'|\sigma^{y}_l)=\frac{\sinh\frac{\gamma(\boldsymbol{\theta})-\gamma(\boldsymbol{\theta'})
 }{2}}{i\sinh\Kr_x^*}\, \mathcal{F}^{(l)}_{m,n}(\boldsymbol{\theta},
 \boldsymbol{\theta}'),\\
 \mathcal{F}^{\mathrm{XY}}_{m,n}(\boldsymbol{\theta},
 \boldsymbol{\theta}'|\sigma^{z}_l)=\frac{\cosh\frac{\gamma(\boldsymbol{\theta})-\gamma(\boldsymbol{\theta'})
 }{2}}{\cosh\Kr_x^*}\, \mathcal{F}^{(l)}_{m,n}(\boldsymbol{\theta},
 \boldsymbol{\theta}'),
 \end{eqnarray}
 where $\mathcal{F}^{(l)}_{m,n}(\boldsymbol{\theta},
 \boldsymbol{\theta}')$ is defined by (\ref{isingff}) and $\gamma(\boldsymbol{\theta})$ by
 (\ref{gammabold}). The same expressions have been
 recently found in \cite{IorgovXY} by the method of separation of variables
 and used to rederive the asymptotics of two-point correlation function in the disordered phase without
 the use of  Toeplitz determinants and Wiener-Hopf factorization method.

 Detailed discussion of the XY chain is beyond the scope of the present paper.
 We would like to mention, however, the works \cite{IKK,KP}, where time- and temperature dependent two-point
 correlation functions of the finite XY chain were expressed in terms of  $2N\times2N$ determinants.
 In fact equivalent representations can be simply deduced from the observation that spin operators, as well
 as the densities  $e^{-\beta H_{\nu}}$,
 are elements of the Clifford group; the entries of the corresponding determinants are
 then given by thermally dressed two-particle form factors. We will report on these issues  in
 a future publication.

 Our last remark concerns the operators appearing in the hamiltonians $H^{\,}_{\varepsilon}$
 and $H'_{\varepsilon}$. It can be checked that they obey Onsager algebra relations for the
 standard generators $\{A_j\}$, $\{G_j\}$:
 \begin{eqnarray*}
 H^{\,}_{\varepsilon}=2\kappa A_1+\mu A_0-2\lambda G_1+\rho A_2,\\
 H'_{\varepsilon}=\mathrm{const}\cdot \left(e^{2\Kr_x}A_1+e^{-2\Kr_x}A_{-1}+2\coth2\Kr_y A_0\right).
 \end{eqnarray*}
 The similarity transformations with $\exp({\Kr A_0})$, $\exp({\Kr'A_1})$ relating these two hamiltonians
 are combinations of elementary Onsager algebra automorphisms described in \cite{Ahn}.

 \ack
 This work was supported by the
  French-Ukrainian program Dnipro M17-2009, the joint PICS project  of CNRS and NASU,
  IRSES project ``Random and Integrable Models in Mathematical Physics'', the Program of Fundamental Research of the Physics and Astronomy
 Division of the NASU and Ukrainian FRSF grants $\Phi$28.2/083 and $\Phi$29.1/028.


\Bibliography{50}
 \bibitem{Ahn}
 Ahn C and Shigemoto K 1991 Onsager algebra and integrable lattice models
 \textit{Mod. Phys. Letts.}~\textbf{A6} 3509--3515
 \bibitem{baxterbook} Baxter R J 1982 \textit{Exactly solved models in statistical mechanics}
 (San Diego: Academic Press)
 \bibitem{BS1} Bazhanov V V and Stroganov Yu G 1985
 Hidden symmetry of the free fermion model: I Triangle equation and
symmetric parametrization
 \textit{Theor. Math. Phys.}~\textbf{62} 253--260
 \bibitem{BS2} Bazhanov V V and Stroganov Yu G 1985
 Hidden symmetry of the free fermion model: II Partition function
 \textit{Theor. Math. Phys.}~\textbf{63} 519--527
  \bibitem{BS3} Bazhanov V V and Stroganov Yu G 1985
 Hidden symmetry of the free fermion model: III Inversion relations
 \textit{Theor. Math. Phys.}~\textbf{63} 604--611
 \bibitem{Bugrij} Bugrij A I 1991 Fermionization of a generalized two-dimensional Ising model
 \textit{Electron-Electron correlation effects in Low-dimensional Conductors and Superconductors} ed
 A A Ovchinnikov and I.~I.~Ukrainskii (Berlin: Springer) pp 135--151
 \bibitem{BIS} Bugrij A I, Iorgov N Z and Shadura V N 2005 Alternative method of calculating the eigenvalues
of the transfer matrix of the $\tau_2$ model for $N=2$ \textit{JETP Letts}~\textbf{82} 311--315
  \bibitem{BL1} Bugrij A I and Lisovyy O 2003 Spin matrix elements in 2D Ising model on the finite lattice
 \textit{Phys. Lett.}~\textbf{A319} 390--394 arXiv:0708.3625 [nlin.SI]
 \bibitem{BL2} Bugrij A I and Lisovyy O 2004 Correlation function of the two-dimensional Ising model on a finite lattice. II \textit{Theor. Math. Phys.}~\textbf{140} 987--1000 arXiv:0708.3643 [nlin.SI]
 \bibitem{FanWu} Fan C and Wu F Y 1970 General lattice model of phase transitions
 \textit{Phys. Rev.}~\textbf{B2} 723--733
      \bibitem{Iorgov1} von Gehlen G, Iorgov N, Pakuliak S, Shadura V and  Tykhyy Yu 2007
 Form-factors in the Baxter-Bazhanov-Stroganov model I: Norms and matrix elements
 \textit{J. Phys.} \textbf{A40} 14117--14138 arXiv:0708.4342 [nlin.SI]
 \bibitem{Iorgov2} von Gehlen G, Iorgov N, Pakuliak S, Shadura V and  Tykhyy Yu 2008
 Form-factors in the Baxter-Bazhanov-Stroganov model II: Ising model on the finite lattice
 \textit{J. Phys.} \textbf{A41} 095003 arXiv:0711.0457 [nlin.SI]
 \bibitem{GreenHurst} Green H S and Hurst C A 1964 \textit{Order-disorder phenomena} (New York: Interscience)
 \bibitem{Hystad} Hystad G 2011 Periodic Ising correlations
 \textit{J. Math. Phys.}~\textbf{52} 013302 arXiv:1011.2223 [math-ph]
  \bibitem{IorgovXY} Iorgov N 2009 Form-factors of the finite quantum XY-chain
  arXiv:0912.4466 [cond-mat.stat-mech]
 \bibitem{IL} Iorgov N and Lisovyy O 2011 Ising correlations and elliptic determinants  DOI:10.1007/s10955-011-0154-6, to appear in \textit{J. Stat. Phys.}
 arXiv:1012.2856 [math-ph]
 \bibitem{IKK} Izergin A G, Kapitonov V S and Kitanine N A 1997 Equal-time temperature correlators
 of the one-dimensional Heisenberg XY chain \textit{Zap. Nauchn. Semin. POMI}~\textbf{245} 173--206
 arXiv:solv-int/9710028
  \bibitem{KP}  Kapitonov V S and  Pronko A G 2003 Time-dependent correlators of local spins
 of the one-dimensional XY Heisenberg chain \textit{J. Math. Sciences}~\textbf{115} 2009--2032
  \bibitem{kaufman} Kaufman B 1949 Crystal statistics. II. Partition function evaluated by spinor analysis
 \textit{Phys. Rev.}~\textbf{76} 1232--1243
 \bibitem{Khachatryan} Khachatryan Sh and Sedrakyan A 2009 Characteristics of 2D lattice models from fermionic realization: Ising and XYZ models
 \textit{Phys.Rev.}~\textbf{B80} 125128	arXiv:0712.0273v2 [cond-mat.str-el]
  \bibitem{OL} Lisovyy O 2006 Transfer matrix eigenvectors of the Baxter-Bazhanov-Stroganov
 $\tau_2$-model for $N=2$ \textit{J. Phys.}~\textbf{A39} 2265--2285
 	arXiv:nlin/0512026 [nlin.SI]
 \bibitem{Onsager} Onsager L 1971 The Ising model in two dimensions \textit{Critical Phenomena in Alloys, Magnets and Superconductors} ed R E Mills, E Ascher and R I Jaffee (New York: McGraw-Hill)
     pp 3--12
  \bibitem{Palmer_Hystad} Palmer J and Hystad G 2010 Spin matrix for the scaled periodic Ising model
 \textit{J. Math. Phys.}~\textbf{51} 123301 arXiv:1008.0352 [nlin.SI]
    \bibitem{smj} Sato M, Miwa T and Jimbo M 1980 Holonomic quantum
 fields V  \textit{Publ. RIMS, Kyoto Univ.}~\textbf{16} 531--584
 \bibitem{Stephen} Stephen M J and Mittag L 1972 A new representation of the solution
 of the Ising model \textit{J. Math. Phys.}~\textbf{13} 1944--1951
 \bibitem{suzuki} Suzuki M 1971 Equivalence of the two-dimensional Ising model to the ground state of the linear XY-model \textit{Phys. Lett.}~\textbf{A34} 94--95
 \endbib
\end{document}